\documentclass{article}

\usepackage{PRIMEarxiv}

\usepackage[utf8]{inputenc} 
\usepackage[T1]{fontenc}    
\usepackage{hyperref}       
\usepackage{url}            
\usepackage{booktabs}       
\usepackage{amsfonts}       
\usepackage{nicefrac}       
\usepackage{microtype}      
\usepackage{lipsum}
\usepackage{fancyhdr}       
\usepackage{graphicx}       
\usepackage[numbers]{natbib}
\graphicspath{{media/}}     

\title{Let's talk about the weather\\ \Large A cluster-based approach to weather forecast accuracy
\thanks{\textit{\underline{Citation}}: 
\textbf{JF Lundell, B Bean, and J Symanzik. Let's talk about the weather: A cluster-based approach to weather forecast accuracy.}} 
}

\author{
 Jill Lundell \\
  Department of Data Science\\
  Dana-Farber Cancer Institute\\
  Department of Biostatistics\\
  Harvard T.H. Chan School of Public Health\\
  Boston, MA 02215 \\
  \texttt{jlundell@ds.dfci.harvard.edu} \\
   \And
 Brennan Bean \\
  Department of Mathematics and Statistics\\
  Utah State University\\
  Logan, UT 84322 \\
  \texttt{brennan.bean@usu.edu} \\
   \And
  J\"{u}rgen Symanzik \\
  Department of Mathematics and Statistics \\
  Utah State University\\
  Logan, UT 84322 \\
  \texttt{Juergen.Symanzik@usu.edu} \\
}

\begin{document}
\maketitle

\begin{abstract}
Improved understanding of characteristics related to weather forecast accuracy in the United States may help meteorologists develop more accurate predictions and may help Americans better interpret their daily weather forecasts. This article examines how spatio-temporal characteristics across the United States relate to forecast accuracy. We cluster the United States into six weather regions based on weather and geographic characteristics and analyze the patterns in forecast accuracy within each weather region. We then explore the relationship between climate characteristics and forecast accuracy within these weather regions. We conclude that patterns in forecast errors are closely related to the unique climates that characterize each region. 
\end{abstract}


From the icy, wet winters along the Great Lakes, to the hot and dry summers in the Southwest, the United States (U.S.) experiences a wide range of climatic extremes. These extremes create unique challenges when forecasting the weather. Understanding forecast errors across such a diverse landscape is equally challenging, requiring multi-dimensional visualizations across space, time, and climate measurements. Better understanding of the nature and patterns in forecast errors across the U.S. helps meteorologists as they strive to improve weather forecasts. It can also help everyday Americans know how much faith to put in the weather forecast on the day of an important event.

The 2018 Data Expo of the Sections on Statistical Computing and Statistical Graphics of the American Statistical Association (ASA) provided an opportunity to explore and compare weather forecast errors across the U.S. Our analysis focused on the question:
\begin{center}
How do weather forecast errors differ across regions of the U.S.?
\end{center} 
This motivating question prompted the subsequent questions:
\begin{itemize}
\item Do U.S. weather stations cluster into regions based on weather characteristics?
\item How do error variables correlate and do these correlations change by region?
\item How do forecast errors change by region and by season? 
\item Where are the best and worst forecast accuracies? 
\item Which variables are important in determining forecast errors?
\end{itemize}
Preliminary results of our analysis are published in the proceedings for the 2018 Joint Statistical Meetings \cite{lundell2018}.

This article is devoted to answering these questions. We use ensemble graphics to create an overall picture of weather forecast errors across different regions of the U.S. \cite{unwin2018ensemble}. Ensemble graphics enhance traditional analyses by connecting several visualizations of the data with adjoining text. This presentation is able to tell a cohesive story of the data more effectively than would be possible with a few disjointed graphics. In Section \ref{clustering}, we summarize the data and then show that the U.S. can be clustered into six well-defined weather regions using the provided climate measurements, elevation, and distance to coast. These clusters, or weather regions, form the basis of our comparison of forecast accuracy across the U.S. through a series of multi-dimensional plots and variable importance analyses described in Section \ref{errorExplore}. in Section \ref{app}, we introduce the interactive application we created to enhance our data explorations. We conclude in Section \ref{conclusions} that the climate differences that distinguish the weather regions of the U.S. also create region-specific patterns and differences in forecast accuracy. Two appendixes are included at the end of this paper to explain data cleaning and how to create the glyphs used in this article.

\section{Weather regions \label{clustering}}

The data contain measurements and forecasts for 113 U.S. weather stations from July 2014 to September 2017. These data can be obtained from our supplemental materials or at the following URL:

\begin{flushleft}
\url{http://community.amstat.org/stat-computing/data-expo/data-expo-2018}. 
\end{flushleft}

Daily measurements for eight different weather metrics were recorded for each location including temperature, precipitation, dew point, humidity, sea level pressure, wind speed, cloud cover, and visibility. Many notable weather events are also textually recorded such as thunderstorms and fog. Daily measurements of the minimum, maximum, and mean were recorded for each metric. Weather characteristics used in this article are listed in Table~\ref{tbl:variables}. Data were supplemented with some geographic information and carefully examined and cleaned. Details on data cleaning, obtaining additional data, and the justification behind our final variable selection are found in Appendix \ref{cleaning}. 

\begin{table}
\caption{List of weather variables included in our analysis. All observations outside the indicated ranges were removed prior to our analysis.}
\label{tbl:variables}       
\begin{tabular}{lcc}
\hline\noalign{\smallskip}
\textbf{Variable} & \textbf{Unit} & \textbf{Range} \\
\noalign{\smallskip}\hline\noalign{\smallskip}
Min/Max Temperature & $^\circ F$ & $[-37, 127]$ \\
Precipitation & in & $[0, 12.95]$ \\
Min/Max Dew Point &  $^\circ F$ & $[-50, 90]$ \\
Min/Max Humidity & \% & $(0, 100]$   \\
Min/Max Sea Level Pressure & inHg & $[28.2, 31.2]$  \\
Mean/Max Wind Speed & mph &  $[0, 70]$\\ 
Min Visibility  & mi &  $[0, 10]$\\
Cloud Cover & okta & \small $\{0, 1,\cdots, 8\}$\\
Distance to Coast & mi & $[0, 807]$ \\
Elevation & ft & $[3, 7422]$\\
\noalign{\smallskip}\hline
\end{tabular}
\end{table}

\subsection{Developing weather clusters \label{cluster}}

The U.S. has been divided into regions based on environmental characteristics such as watersheds and climate \cite{ecoregions}\cite{Briggs}. We examined the set of existing environmental regions and were unable to find one that made sense in terms of weather in the context of this analysis. We created our own weather regions by clustering the weather stations based on the metrics in Table~\ref{tbl:variables}. Thus, clusters are defined by weather characteristics observed at each station. We use these clusters to determine how weather forecast error patterns are related to the unique climate measurements of a particular region. A review of existing weather regions and how they correspond to our weather regions is discussed in Section \ref{historic}. Data were aggregated across each weather station by taking the mean and standard deviation of each variable in Table~\ref{tbl:variables} for each of the 113 weather stations over the period of record.  

Hierarchical clustering \cite{Friedman2001} with Euclidean distance and Ward's minimum variance clustering method \cite{Murtagh2014} was used to identify clusters. The clusters were examined spatially to determine the performance of the clustering method and select the final number of clusters. We wanted to ensure the weather station clusters were of a sufficient size to be practical. Five clusters resulted in one cluster that included all of the stations from the Midwest to the East Coast which we think is too large because of the differences in coastal and inland climates. Seven clusters produced a cluster that contained only five weather stations which is too small. Thus, we chose six clusters to divide the U.S. into weather regions. 

\begin{figure}
\begin{center}
\includegraphics[width=1\linewidth]{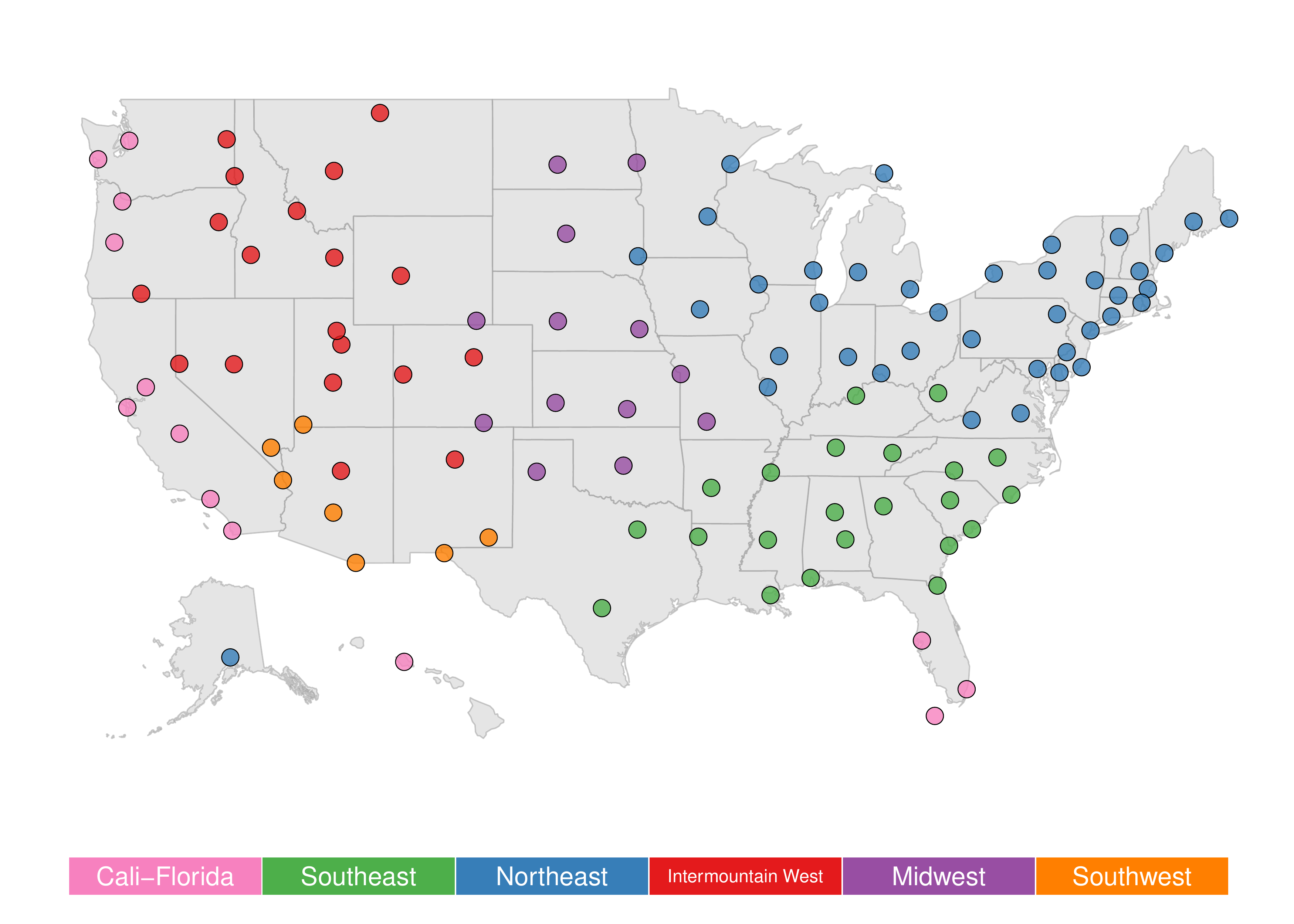}
\caption
{Map of the six weather regions. The color band at the bottom identifies each region by name and color}
\label{fig:clust}
\end{center}
\end{figure}

\begin{figure}
\begin{center}
\includegraphics[width=0.5\textwidth, trim = 0cm 2cm 0cm 2cm]{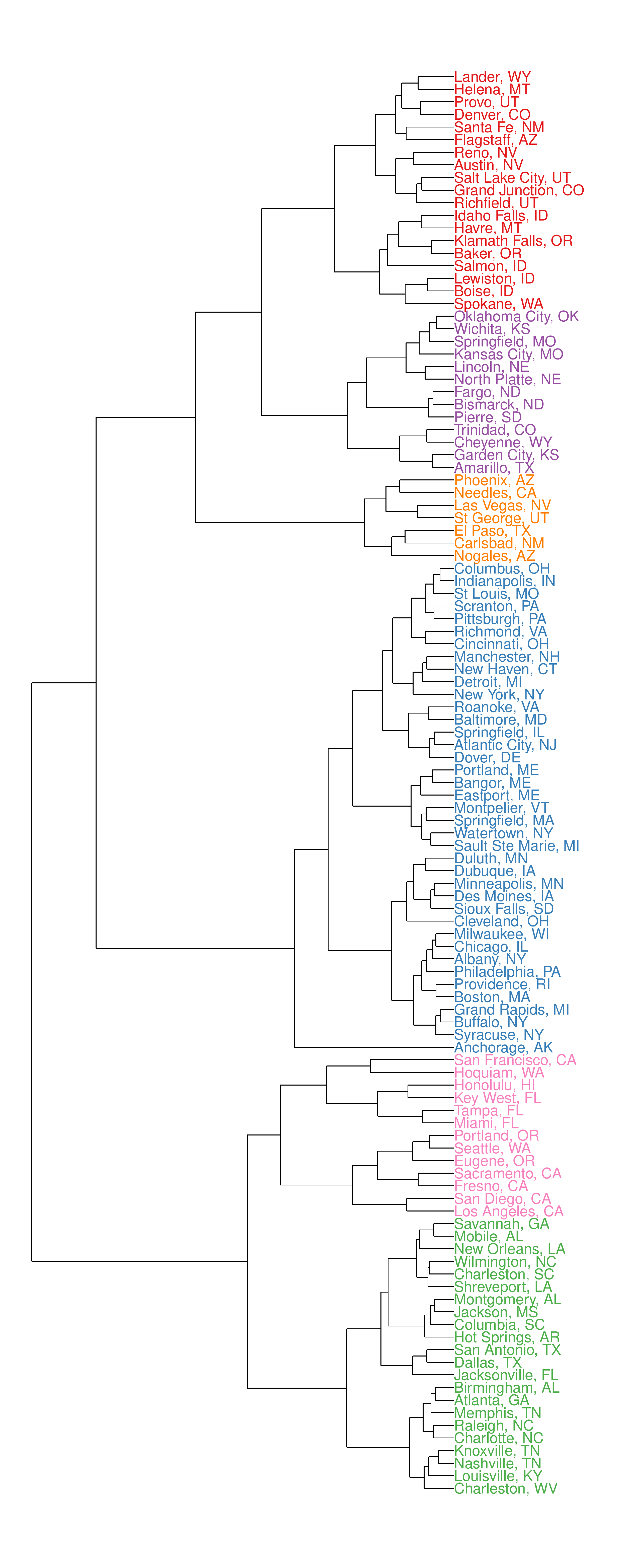}
\includegraphics[width=\textwidth]{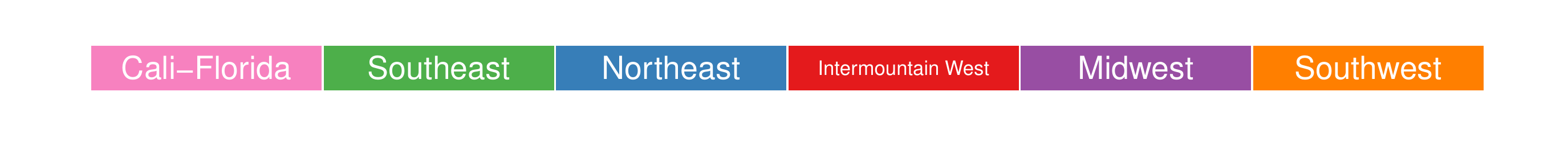}
\caption
{Dendrogram of weather clusters identified in Figure \ref{fig:clust}}
\label{fig:dendro}
\end{center}
\end{figure}

\begin{figure}
\begin{center}
\includegraphics[width=1\linewidth]{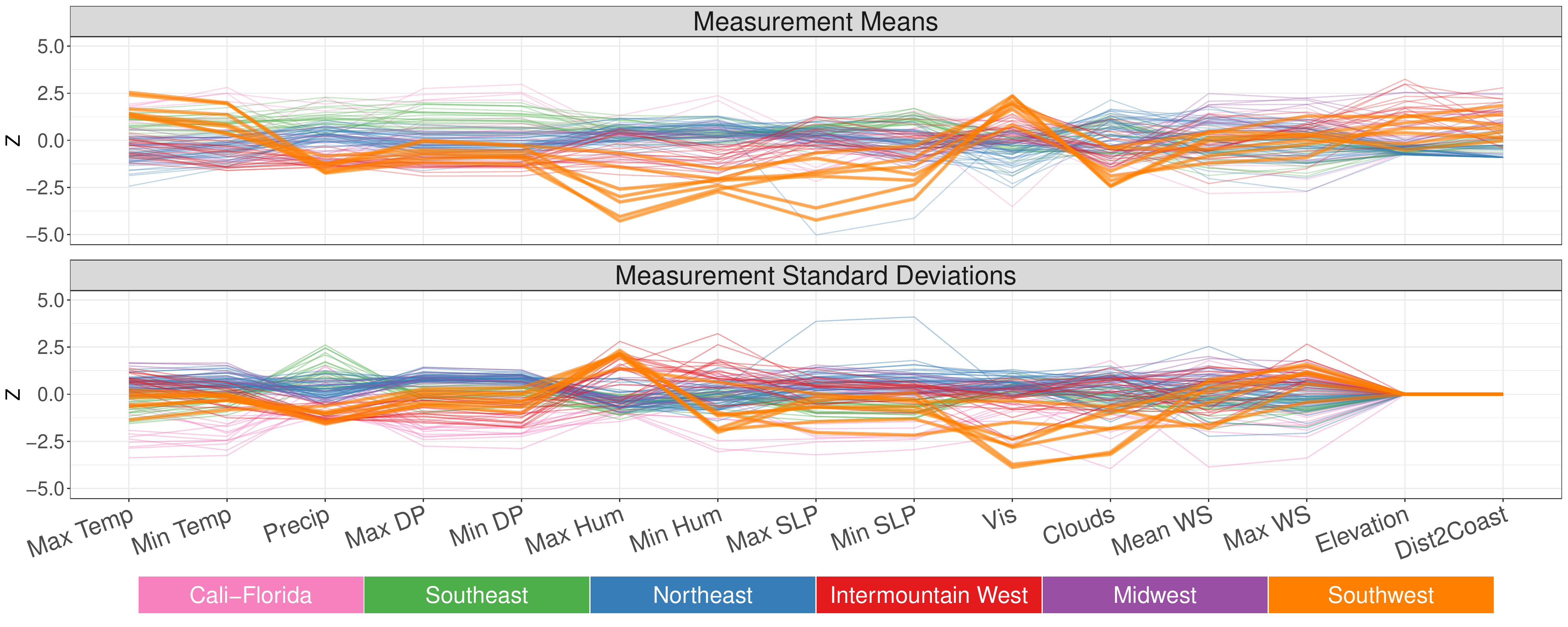}
\caption{Parallel coordinate plot of the means and standard deviations of the weather variables listed in Table~\ref{tbl:variables}. Each line in the plot represents one of the 113 weather stations. The color of the lines match the weather region to which the station belongs. An interactive app is available that allows for better identification of regional trends.The Southwest region is highlighted in this graph to emphasize its weather characteristics}
\label{fig:pcp}
\end{center}
\end{figure}

Figures~\ref{fig:clust} and \ref{fig:dendro} show the results of the cluster analysis. Figure~\ref{fig:pcp} shows a parallel coordinate plot of the characteristics for each weather region. The Z-score for mean and standard deviation for each of the variables in Table~\ref{tbl:variables} was computed and plotted on the parallel coordinate plot. It is difficult to distinguish the six weather regions from each other so an interactive app was created that provides a better view of the features of each cluster. The app is discussed in Section \ref{app}.

The names and characteristics of each weather cluster are as follows: 
\begin{itemize}
\item \textbf{Cali-Florida} (13 stations): Warm and humid with high dew point and pressure. Low variability in almost all measurements.   
\item \textbf{Southeast} (22 stations): Warm and humid with lots of rain. High variability in precipitation and low variability in temperature.
\item \textbf{Northeast} (39 stations): Cold, humid, and low visibility. High variability in temperature, dew point, and pressure. 
\item \textbf{Intermountain West} (19 stations): Cold and dry, with high variability in temperature, wind speed, and pressure. Low variability in precipitation and dew point.  
\item \textbf{Midwest} (13 stations): Landlocked with high wind speed and high variability in temperature, pressure, and wind speed. 
\item \textbf{Southwest} (7 stations):  Warm, sunny, and dry with little variation in temperature or precipitation. High variability in wind speed and humidity. 
\end{itemize}

\subsection{Comparison to existing climate regions \label{historic}}

Ecological and climate regions have been developed for the U.S. in other studies. Many of these studies focused on smaller regions in the U.S., but a few have looked at the U.S. as a whole. Clustering methods and the variables used to identify clusters differ from study to study. The ecological regions of North America defined by the Commission for Environmental Cooperation \cite{ecoregions} used ecosystems to develop regions. Air, water, land, and biota, including humans, were used to create the ecoregions. These ecoregions show a strong longitudinal trend that corresponds well with the longitudinal trends in our clusters. Clusters were not determined by statistical clustering methods, but by careful assessment of ecological properties across North America. 

The National Oceanic and Atmospheric Administration (NOAA) developed climate regions that incorporate seasonal temperature and precipitation information \cite{karl1984regional}. These regions differ substantially from the North American ecological regions as they also have a lateral trend in addition to the longitudinal trend and are constrained by state boundaries.  Spectral curves assessing drought and wet spells were used to define the NOAA regions \cite{diaz1983drought}. The NOAA regions correspond roughly to our general weather regions despite region borders being defined by state boundaries.  The north/south division in the eastern U.S. closely aligns with our cluster division in that area. The major east/west division in our clusters is in a similar location to the NOAA clusters as well. 

The International Energy Conservation Code (IECC) climate clustering of the U.S. \cite{Briggs} and subsequent reclassification by Hathaway, et al. \cite{hathaway2013statistical} divided the U.S. into fourteen regions based on temperature, dew point, wind speed, and radiation. Cluster methods included K-means clustering and Monte-Carlo sifting. Both sets of regions show a strong lateral trend in the Eastern U.S. These regions also show distinct separation of the West coast and Southwest deserts from the rest of the Western U.S. Similar trends are also seen in our clusters. The lateral trend in the Eastern U.S. is not as strong in our clusters, but this is likely because we chose a smaller number of weather clusters. The inclusion of additional variables insensitive to lateral trends such as distance to coast, elevation, and humidity, all serve to reduce the lateral separation in our clusters.

One key difference between our weather regions and the regions seen in other studies is that we combine Florida and the Pacific coast into a single weather region. This is likely a result of our choice to omit geographic proximity of weather stations in the cluster analysis calculations and consider only similarities in weather patterns. Both Florida and the Pacific coast experience less seasonality in their weather patterns than the rest of the country. This results in smaller than average standard deviations for many of the climate variables in both of these regions. These small standard deviations create a measure of closeness between Florida and the Pacific coast which likely explains why these two geographic areas all into a single cluster when working with six or fewer clusters. The Florida and Pacific stations separate into separate clusters when using seven clusters with exception of two stations from the Pacific Coast that cluster with the Florida stations. Hawaii and Alaska are either ignored in the literature or placed in their own regions. Because we did not use spatial proximity as a clustering variable and we assigned all weather stations to one of our six weather clusters, Hawaii and Alaska are clustered with Cali-Florida and the Northeast respectively. Our clusters show that weather patterns typically have strong spatial correlations, with temperate coastal regions being a notable exception.

\section{Forecast error explorations \label{errorExplore}}

Given the clear separation of the country into distinct weather regions, we seek to determine if there are clear differences in forecast error patterns among the regions. Forecasts were restricted to minimum temperature, maximum temperature, and the probability of precipitation. The forecast error for minimum and maximum temperature is calculated as the absolute difference between forecast and measurement. The forecast error for precipitation is measured using the Brier Skill Score (BSS), a well-known measure of probabilistic forecast accuracy \cite{Weigel2007}. It is defined for a particular weather station as  
\begin{equation}
\mbox{BSS} = 1 - \frac{\sum\limits_{i=1}^N\sum\limits_{j = 0}^M\left(Y_{ij} - O_{i}\right)^2}{\sum\limits_{i=1}^N\sum\limits_{j = 0}^M\left(P - O_{i}\right)^2}
\end{equation}
where
\begin{itemize} 
\item $Y_{ij} \in \left[0, 1\right]$ is the predicted probability of rain on day $i$ with forecast lag $j$;
\item $O_{i} \in \left\{0,1\right\}$ is a binary variable with value $1$ if \textit{any} precipitation fell during the day and 0 otherwise. We define a precipitation event as a positive precipitation measurement or the inclusion of the words ``rain" or ``snow" in the event information;
\item $P \in \left[0,1\right]$ is the average daily chance of precipitation over the period of interest, defined as $P = \frac{1}{N}\sum\limits_{i=1}^NO_i$;
\item $N$ denotes the number of days of recorded precipitation in the period of record and $M \in \left\{0, \ldots, 5\right\}$ denotes the number of forecast lags.
\end{itemize}

Note that the $\mbox{BSS} \in (-\infty, 1]$, with 1 indicating a perfect forecast skill and movement towards $-\infty$ indicating worse forecasts. We chose to use $1-\mbox{BSS}$ so all three error variables are consistent in orientation. The following subsections explore differences in forecast errors both between and within the previously defined weather regions visualized in Figure \ref{fig:clust}. Forecast errors are averaged over lag and in some cases averaged over month in each graph. The visualizations in the following subsections confirm our hypothesis that different weather regions experience distinctly different weather forecast error patterns.  

\subsection{Error correlations}

\begin{figure}
\includegraphics[width = \textwidth]{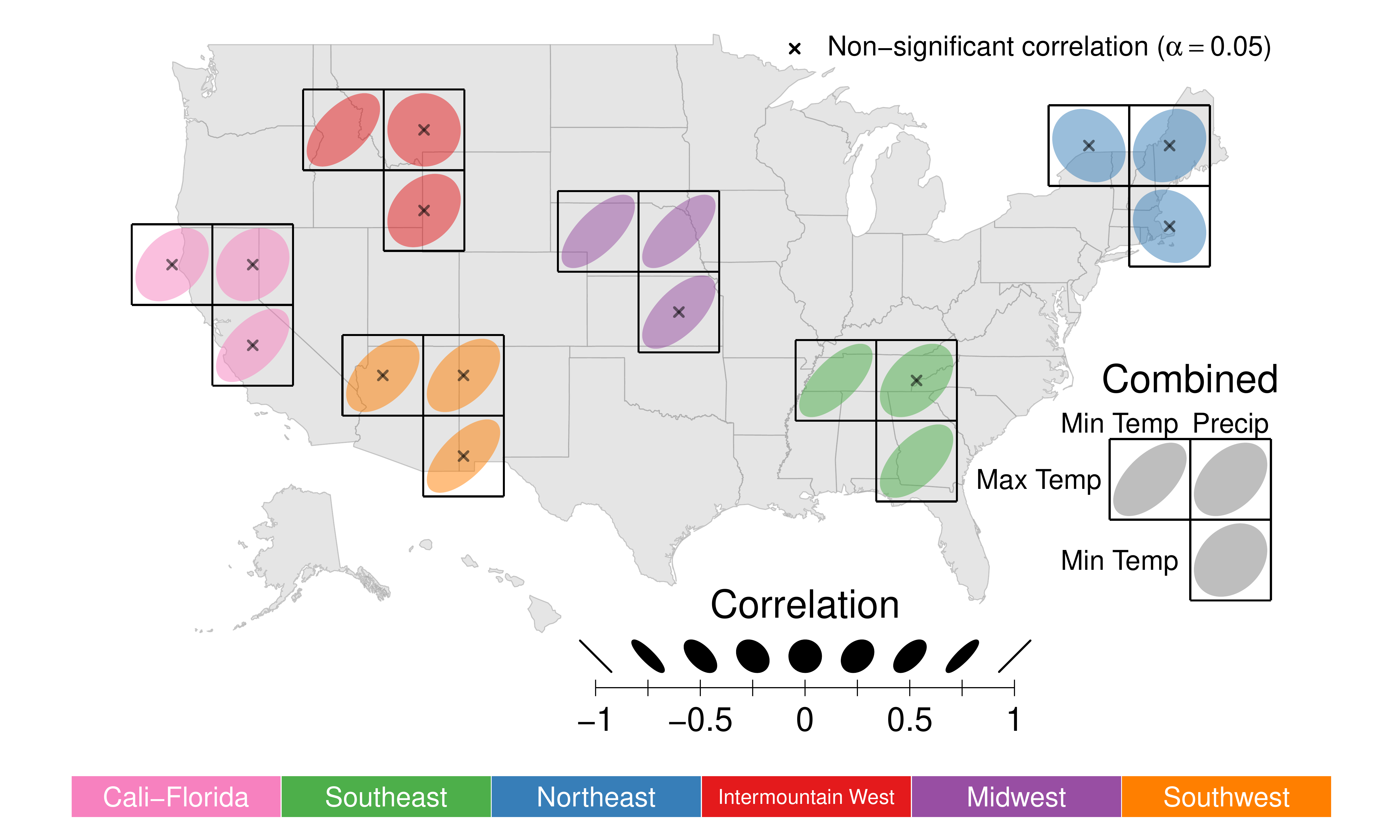}
\caption{Spearman correlations between forecast error variables represented as ellipses superimposed on a map of the United States. The p-value for each correlation is compared against a 0.05 level of significance}
\label{fig:corGlyph}
\end{figure}

Are the forecast errors for the three different measurements (i.e., minimum temperature, maximum temperature, and precipitation) correlated with each other? How do these relationships change between the different weather regions? We explore such correlations through the use of correlation ellipses \cite{Murdoch1996} superimposed on a map of the U.S. in Figure \ref{fig:corGlyph}. We calculated Spearman correlations between each pair of measurements for the locations within each cluster. The sign of the correlation coefficient is denoted by the slope of the ellipse and the strength of correlation is denoted by the width of the ellipse. 

All of the correlations between error variables are positive except for correlations between minimum temperature and the other two variables in the Northeast. The strongest relationships are seen in the Midwest, the South and the Southwest. The weakest relationships are found in the Northeast. Only a few cluster-specific correlations are significant. This is likely due to the small number of stations in many of the weather regions. However, the overall correlations for the 113 weather stations are all positive and significant. This indicates that areas with good predictions for one forecast variable have generally good predictions for the other forecast variables as well. The weakest correlations are between minimum temperature and precipitation predictions. Although there are relationships between the three weather forecast variables, those relationships are not particularly strong and the strength differs within each region. The observations made using this correlation ellipse map illustrate how this plot style facilitates multi-dimensional comparisons across space. Information on the calculations and implementation of the correlation glyphs can be found in Appendix~\ref{polar}. 

\subsection{Error scatterplots\label{correlations}}

\begin{figure}
\begin{center}
\includegraphics[width=.52\linewidth, trim = {0 1cm 0 0}]{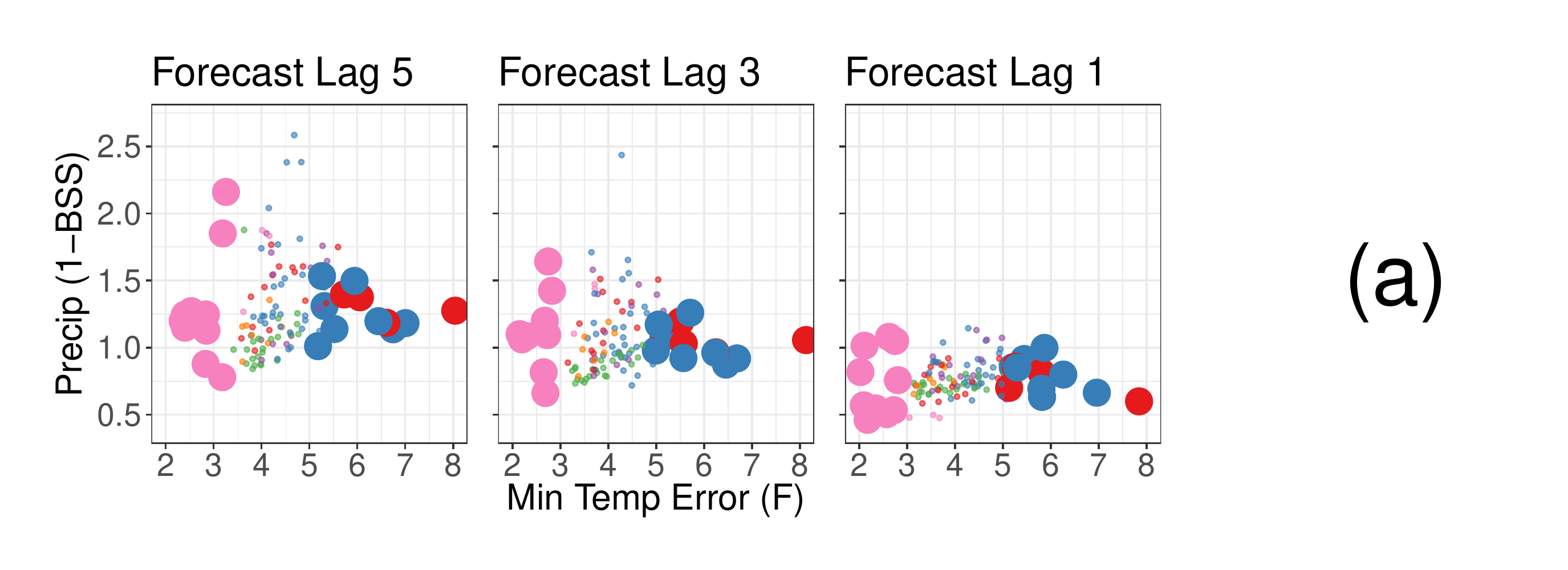}
\includegraphics[width=.52\linewidth, trim = {0 0 0 1cm}]{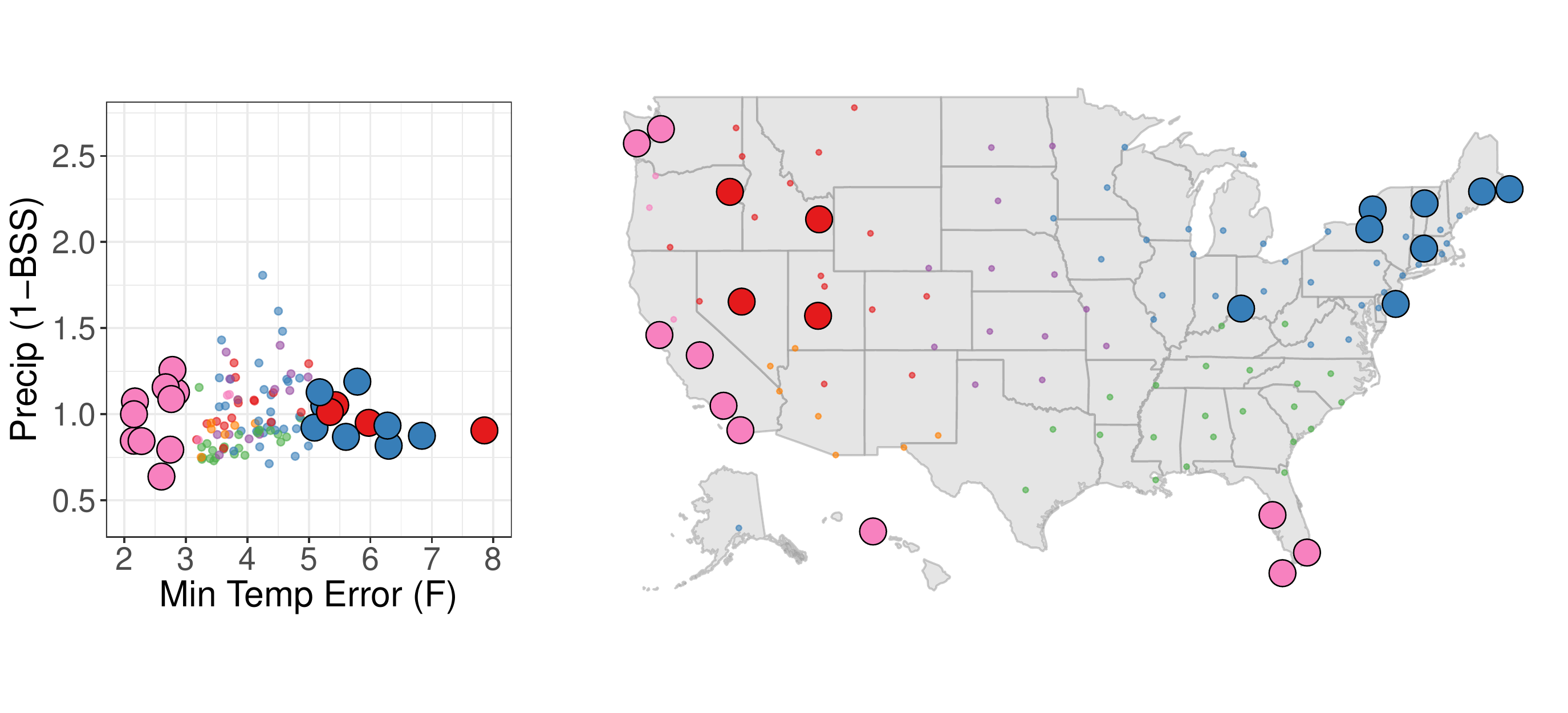}
\includegraphics[width=.52\linewidth, trim = {0 1cm 0 0}]{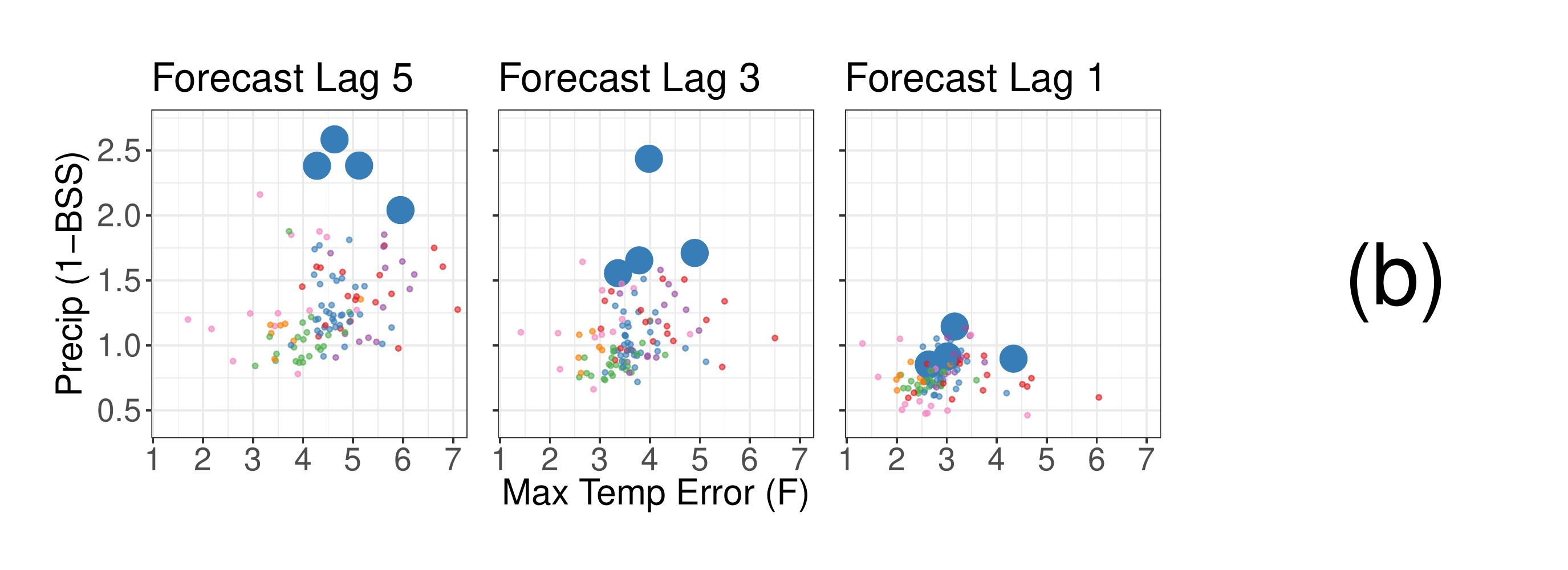}
\includegraphics[width=.52\linewidth, trim = {0 0 0 1cm}]{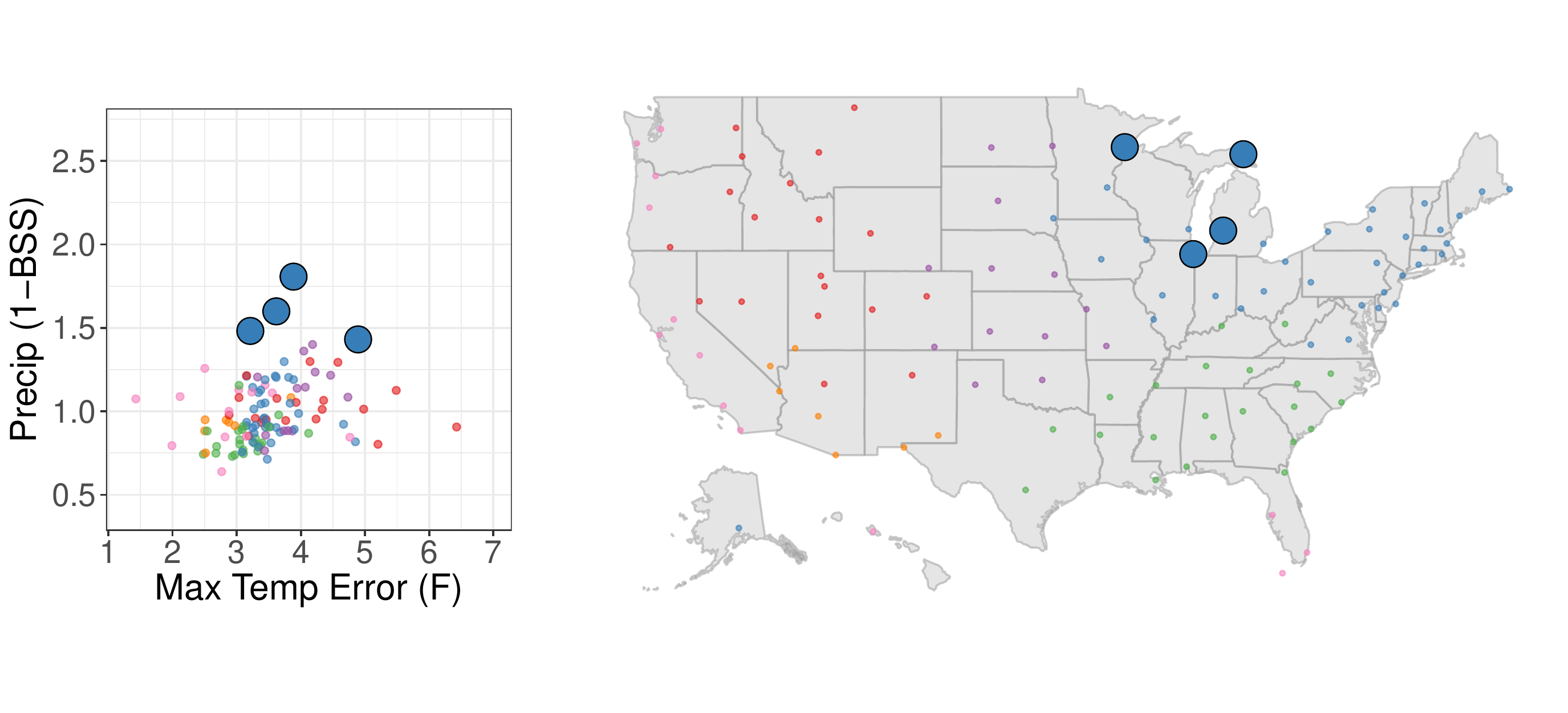}
\includegraphics[width=.52\linewidth, trim = {0 1cm 0 0}]{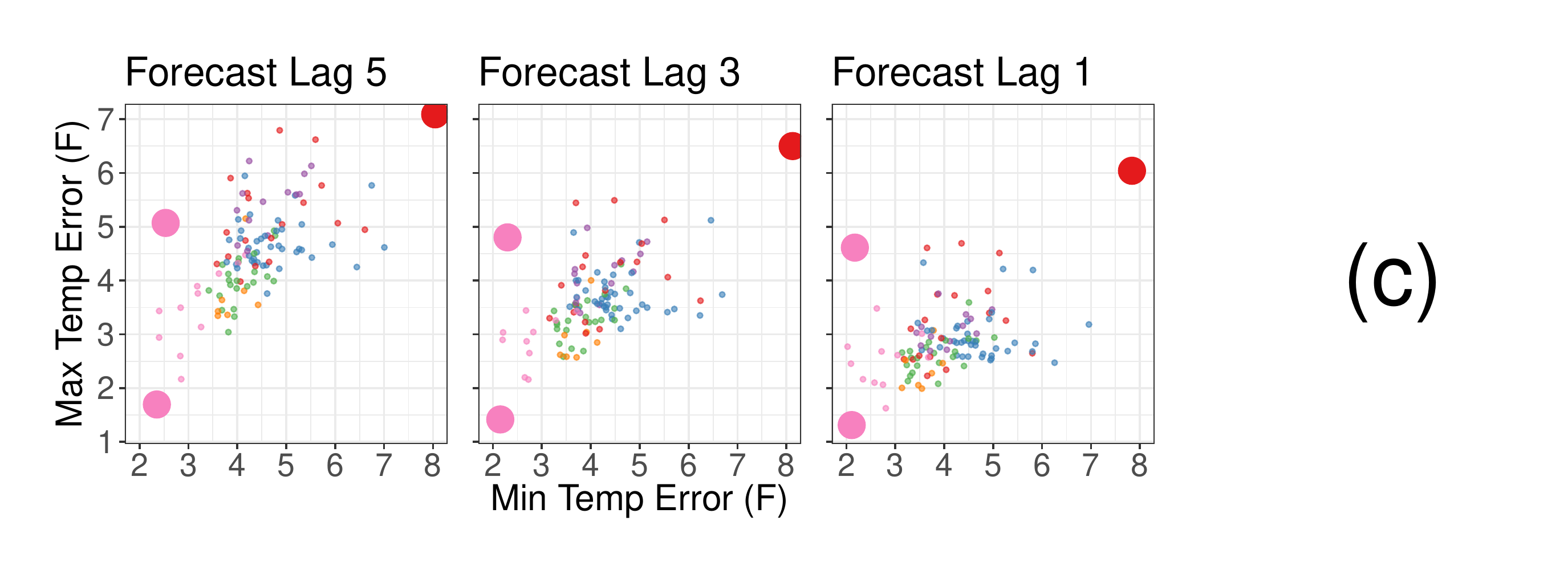}
\includegraphics[width=.52\linewidth, trim = {0 0 0 1cm}]{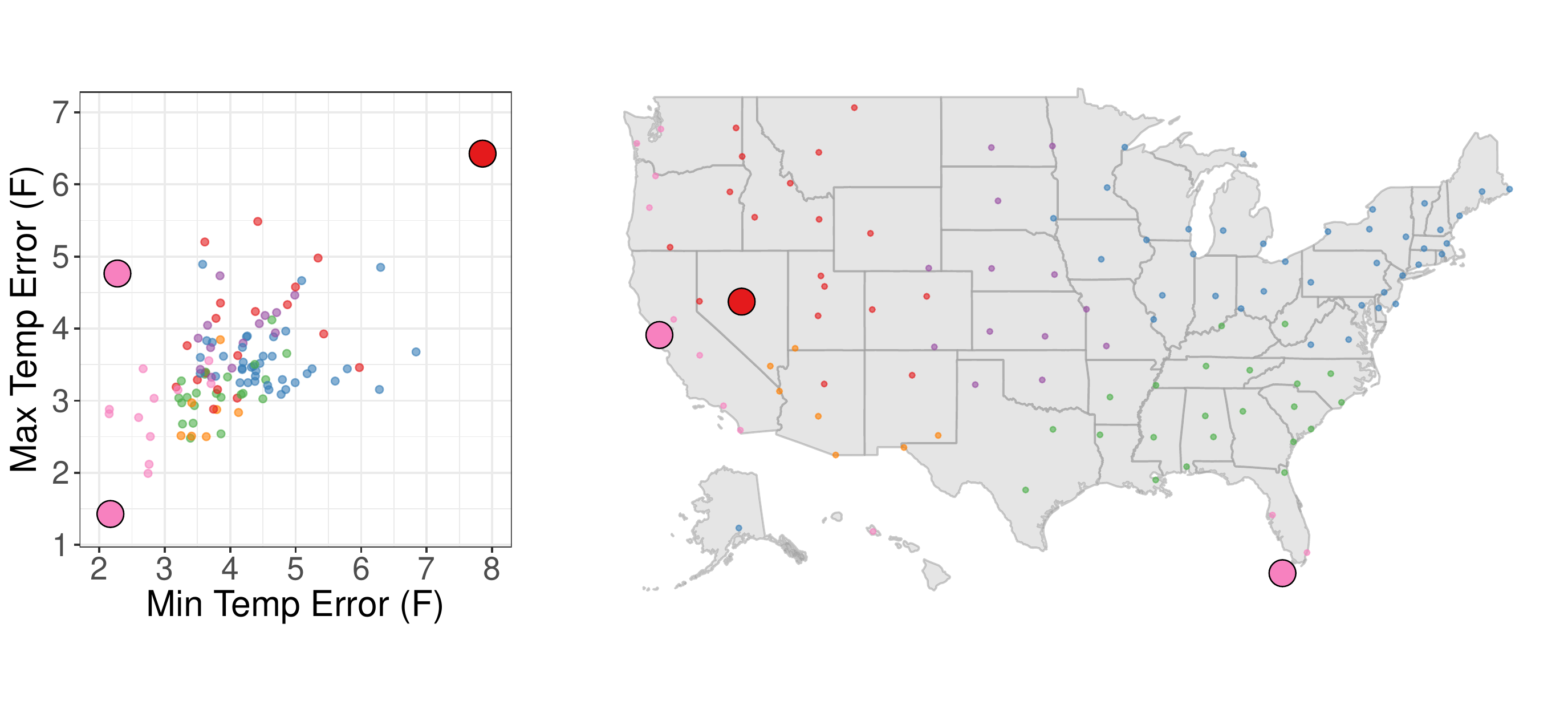}
\includegraphics[width=\linewidth]{ScatterLegend.pdf}
\caption{Scatterplots comparing the three forecast error variables. The scatterplot to the left of the map is aggregated over all forecast lags. Points of interest discussed in the text are highlighted in the respective plots}
\label{fig:scatALL}
\end{center}
\end{figure}

Scatterplots reveal outliers and overall trends within weather regions and across forecast lag. Forecast lag is defined as the number of days between the day of forecast and the day being forecast. Thus, same day forecasts would have a lag of 0, one day prior forecasts a lag of 1, and so on. Because we are comparing three variables spatially and temporally across the U.S., static graphs are not optimal for assessing all relationships of interest. We constructed an interactive scatterplot app that facilitates examination of trends between the three forecast error variables aggregated across all forecast lags or for individual forecast lags. Figure~\ref{fig:scatALL} (a-c) shows examples of plots from the interactive app. The figure shows the scatterplot for the data aggregated over all forecast lags, as well as the scatterplots for lags of 5, 3, and 1, to illustrate how forecast accuracy changes over forecast lag. 

Figure~\ref{fig:scatALL}(a) compares minimum temperature forecast accuracy with precipitation accuracy. Weather stations with the worst predictions of minimum temperature are located in New England and the Intermountain West. New England is known for extreme winter weather and the frequency of extreme weather events seems to be increasing \cite{Cohen2018}. This likely contributes to the struggle these stations have predicting minimum temperature. The worst predictor of minimum temperature is Austin, Nevada. This location is addressed further in Figure~\ref{fig:scatALL}(c). Cali-Florida uniformly has the best predictions of minimum temperature. However, Cali-Florida also has some of the greatest variability in precipitation prediction accuracy when examining individual lags. 

Figure~\ref{fig:scatALL}(b) compares maximum temperature prediction accuracy with precipitation accuracy. Four weather stations in the Great Lakes region have the worst precipitation predictions in the dataset. Poor precipitation forecast accuracy in this region illustrates the difficulty in forecasting lake-effect snow. This phenomenon is discussed in greater depth in Section~\ref{glyphs}. Precipitation forecast accuracy for the Great Lakes region improves substantially as the forecast lag decreases and forecasts with lag 1 are as accurate as the rest of the nation.  

Figure~\ref{fig:scatALL}(c) shows the relationship between minimum and maximum temperature forecast accuracy. Three outliers stand out in these scatterplots, namely Key West, Florida, Austin, Nevada, and San Francisco, California. Key West predicts both minimum and maximum temperature more accurately then any other weather station. Key West also ranks in the top five for lowest variability in eight of the weather variables, which likely explains the accurate forecasts. Austin is the poorest predictor of both measures. Seventy miles along the ``loneliest highway in America" \cite{Austin} separate Austin from its weather measurements which were collected in Eureka, Nevada. The poor predictions for maximum and minimum temperature can be explained by the change in climate over such a large distance. This is reflected in a negative prediction bias of around 5$^\circ$F for maximum temperature and a positive bias of around 7$^\circ$F for minimum temperature. San Francisco has good predictions of minimum temperature and poor predictions for maximum temperature. This phenomenon is further explained in Section~\ref{glyphs}. 

The interactive app developed in conjunction with this project allows for further investigation of forecast accuracy trends. The app is discussed in Section \ref{app}.

\subsection{Seasonal trends\label{glyphs}}

The position of the U.S. in the northern hemisphere makes most of the country subject to distinct weather seasons. Seasons are most pronounced in the northern U.S. We hypothesize that the forecast error behavior is inextricably linked to this seasonality. We explore this through a series of space-time graphs. Modeling space and time simultaneously creates a three-dimensional problem usually visualized as small multiples. Small multiples are ``a series of graphics, showing the same combination of variables [e.g., latitude and longitude], indexed by changes in another variable [e.g., time]" \cite{Tufte2002}. The issue with this approach is that it becomes difficult to visually comprehend all but the most drastic changes from graph to graph. One alternative that allows simultaneous visualizations of both space and time is through the use of glyphs, or symbols, that allow for multi-dimensional visualizations in a spatial context \cite{Carr1992}\cite{Wickham2012}.

\begin{figure}
\includegraphics[width = 0.85\textwidth]{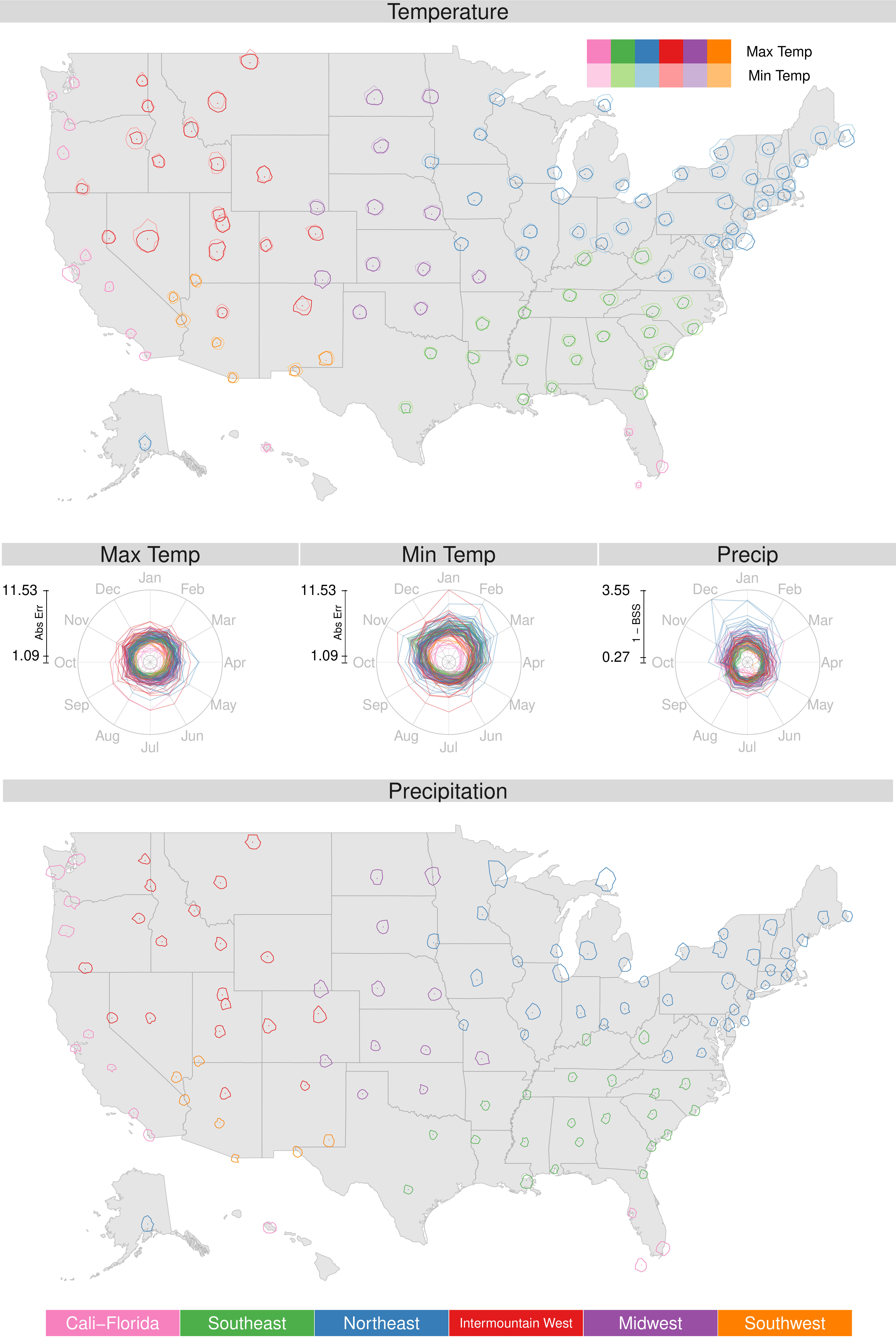}
\caption{Glyph plots of weather forecast accuracy averaged by month. The error is represented as the scaled distance from a center point to the edge of a polygon beginning with January at the 12:00 position and proceeding clockwise}
\label{fig:glyphPlot}
\end{figure}

Figure \ref{fig:glyphPlot} shows glyph plots of seasonal forecast errors throughout time. The forecast error is visualized as the scaled distance from a center point to the edge of a polygon with twelve observations starting with January at the 12:00 position and proceeding clockwise. The asymmetry of the glyphs about their center points illustrates how forecast errors change across time and across space. For example, locations in the Northeast are worse at forecasting precipitation in the winter than in the summer, while locations in the Southeast forecast precipitation equally well throughout the year. 

In addition to highlighting forecasting asymmetries, Figure \ref{fig:glyphPlot} reveals location-specific anomalies. For example, San Francisco, California, predicts minimum temperatures well all year, but only predicts maximum temperatures well in the winter months. This is likely due to chilling coastal fogs known to frequent the region throughout the year that can create sharp temperature differences over short distances \cite{Nolte2016}. The struggle to predict temperature seems reasonable in light of these facts as this measurement location is more than 11 miles inland from the forecast location. The issue is likely less pronounced in the winter because the contrast between inland and coastal temperatures is reduced.

Maximum temperature predictions are particularly poor in the summer months in Austin, Nevada. It is unclear why predictions are worse in the summer than in the winter.  

Another location-specific anomaly of note is the drastic seasonality of precipitation forecasts for locations surrounding the Great Lakes, as observed in Figure \ref{fig:glyphPlot}. The error scatterplots in Figure~\ref{fig:scatALL}(b) show that precipitation accuracy is poor in this region, but the seasonality of the predictions cannot be observed in the scatterplots. The unusually bad forecasting in the winter is likely due to lake-effect snow which is prevalent in the region. Up to 100\% more snow falls downwind of Lake Superior in the winter than would be expected without the lake-effect \cite{Scott1996}. This area has been previously identified as having the most unpredictable precipitation patterns in the nation \cite{Silver2014}. The above examples demonstrate the ease with which comparisons can be made across space and time with these glyph-based plots. Information about how to generate the glyphs is included in Appendix~\ref{polar}.

\subsection{Variable importance\label{importance}}

\begin{figure}
\begin{center}
\includegraphics[width=1\linewidth]{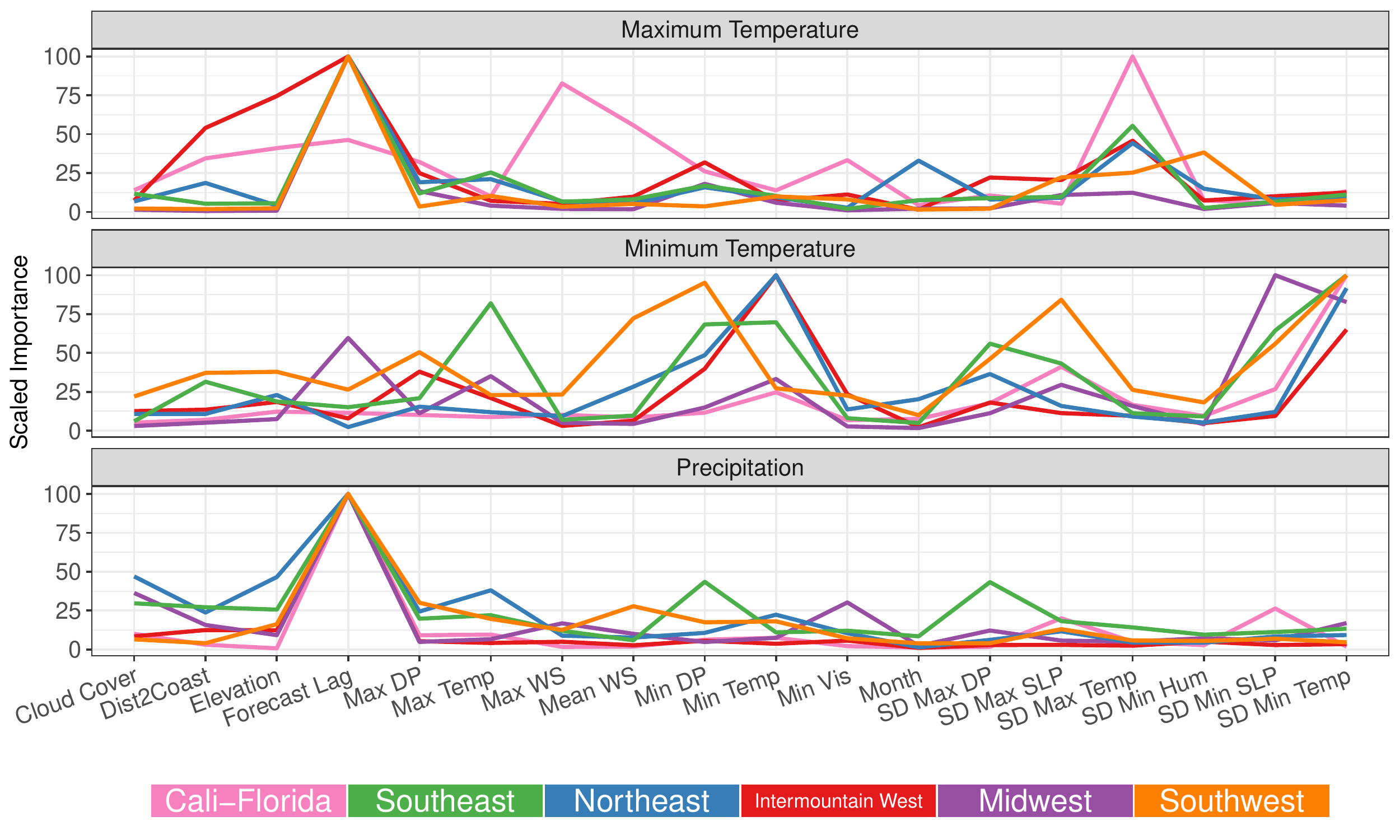}
\caption{Variable importance for each of the three forecast accuracy measurements. Variable importance measures have been rescaled to make the measures directly comparable between weather regions and accuracy measures }
\label{fig:vi}
\end{center}
\end{figure}

The differences in forecast error patterns across regions prompt identification of the most important climate measurements for predicting forecast error. We used random forests \cite{breiman2001random} to determine which weather variables had the greatest impact on the forecast errors. The data were aggregated over forecast lag and month. Three random forest models were generated for each weather region using the forecast error variables as the response. The means and standard deviations for each of the weather variables listed in Table~\ref{tbl:variables} and the forecast lag were the predictor variables. Figure~\ref{fig:vi} contains three parallel coordinate plots that show the variable importance measures in each region for each forecast error variable. 
The importance measures obtained from random forests were recentered by subtracting the minimum importance measure and then rescaled to the interval (0, 100) by dividing by the maximum importance measure of the recentered values for each weather cluster and forecast error variable combination, and finally multiplying by 100. Thus, the most important variable within each weather region has a value of 100 and the least important has a value of 0 for each error measure. This allows direct comparisons of importance between weather regions and across error measures. 

Figure~\ref{fig:vi} shows that the most important variable for the precipitation error is forecast lag regardless of weather region. None of the other variables are very important relative to lag. The Southeast shows minimum dew point (DP) and the standard deviation of maximum dew point as being somewhat important. Cloud cover is important for the precipitation error in the Northeast. 

Forecast lag is also the most important variable for the maximum temperature error for all weather regions except Cali-Florida. The standard deviation of maximum temperature and maximum wind speed (WS) are more important than lag in Cali-Florida. The variability in maximum temperature is also important for the Southeast, Northeast, and the Intermountain West. Distance to coast (Dist2Coast) and elevation are important for the maximum temperature error in the Intermountain West.  

Variables that are important for the minimum temperature error varied substantially across weather regions. The variability in minimum temperatures is important for all regions, but other important variables differ widely from region to region. Minimum temperature is the most important for the Northeast and Intermountain West, but maximum temperature is important for the Southeast. Minimum dew point and the variability in the maximum sea level pressure (SLP) are important in the Southwest while variability in minimum sea level pressure is the most important for the Midwest, Southeast, and Southwest. Forecast lag is not particularly important for any of the regions except for the Midwest.  

\section{Interactive application \label{app}}

It is difficult to identify the patterns in climate measurements and forecast errors for all weather regions with static visualizations. We developed an interactive Shiny app to enhance our weather data explorations. This app can be accessed at \begin{center}\url{https://jilllundell.shinyapps.io/finaldataexpoapp/}.\end{center} The first tab of the app is an interactive version of the parallel coordinate plot introduced in Figure~\ref{fig:pcp}. The app allows the user to select a weather region which is highlighted on the graph. Characteristics of the selected region can be easily seen and compared to all other observations. 
%
%

The second tab of the app is an interactive scatterplot. Figure \ref{fig:scatALL} (a-c) shows examples of the graphs generated in this tab. The user can select up to two of the three forecast error variables to be on the axes. The forecast lag can also be selected. Points on the scatterplot can be brushed or clicked and the selected points show up on a map of the U.S. Information about selected stations is listed in a table under the graph. The idea of linked brushing between scatterplots and maps was first introduced in Monmonier \cite{brush}. This app allows for a more complete exploration of outliers and trends in the data across forecast lags and between error variables than a static graph.

\section{Conclusions\label{conclusions}}

Climate patterns in the United States cleanly separate into six recognizable regions through a cluster analysis using the means and standard deviations of the weather variables provided in Table \ref{tbl:variables}. We explored the relationship between the three weather forecast variables (i.e., minimum temperature, maximum temperature, and precipitation) using correlation ellipses shown in Figure \ref{fig:corGlyph}. We found that all clusters show signs of positive correlations among the error variables with the exception of the Northeast cluster. 

We visualized the pairwise relationship between forecast errors through a series of scatterplots across all forecast lags in Figure \ref{fig:scatALL}. These plots highlight the superiority of locations in the Cali-Florida region for predicting minimum temperature across all lags, and also show that the poor precipitation predictions of the Great Lakes region are mostly confined to forecasts greater than lag 2. Lastly, the abnormally high errors in Austin, Nevada, are likely a product of the large distance between forecast and measurement locations.  

We explored seasonal differences of forecast errors in Figure \ref{fig:glyphPlot} and observed that seasonal differences in forecast errors tend to be more pronounced in northern, inland clusters than southern clusters. We also showed that location specific anomalies, such as the asymmetry in seasonal maximum temperature forecast errors in San Francisco and the precipitation forecast errors near the Great Lakes, have plausible explanations in the literature.

Next, we compared the important variables in determining forecast errors across clusters using scaled random forest variable importance measures in Figure \ref{fig:vi}. These measures demonstrate that forecast lag is most important in determining the maximum temperature and the precipitation forecast errors, but not important in predicting the minimum temperature forecast errors. Many clusters place similar importance on a few variables, but there are some variables that are important only in a single cluster, such as the importance of maximum wind speed in predicting the maximum temperature forecast error in Cali-Florida. 

For further insight regarding the nature of forecast errors across these six clusters, we refer readers to our R shiny app described in the previous section. A current version of the app can be found at the following URL:
\begin{center}
\url{https://jilllundell.shinyapps.io/finaldataexpoapp/}
\end{center}

This app, in conjunction with the visualizations presented in this article, reinforces the idea that the U.S. cleanly clusters into well defined weather regions and patterns in forecast errors are closely related to the unique climates that characterize each region.

The visualizations in this paper, both interactive and static, were designed to be scalable for larger weather datasets. We anticipate illustrating this capability on an expanded set of stations in the future. An expanded analyses will also serve to validate the regional patterns observed and described in this paper. In addition, we anticipate adapting several of the static glyph plots presented in this paper for interactive use. Greater interactivity will allow for more detailed explorations of weather patterns in the United States across both time and space. 

\section{Acknowledgements}
The authors would like to thank the Sections on Statistical Computing and Statistical Graphics of the ASA for providing the data used in this analysis. The primary analytical tool for this analysis was the R \cite{R}. Additional data information regarding specific measurement locations were provided in the weatherData R package \cite{weatherData}. Distance and spatial calculations made use of the fields \cite{fields}, geosphere \cite{geosphere}, mapproj \cite{mapproj}, rgdal \cite{rgdal}, and sp \cite{sp} R packages. Other data manipulations and visualizations made use of the tidyverse \cite{tidyverse}, as well as the ggforce \cite{ggforce}, latex2exp \cite{latex2exp}, RColorBrewer \cite{RColorBrewer}, and reshape2 \cite{reshape2} R packages. Variable importance models made use of the randomForest \cite{randomForest} R package.  also

\appendix 

\section{Data cleaning}
\label{cleaning}

We primarily used the dataset provided by the Data Expo to perform the analyses described in the article. We supplemented the provided location information with elevation and distance to the nearest major coast. Elevation information was obtained for each location through Google's API server \cite{GoogleAPI} via the rgbif R package \cite{rgbif}. Distance to coast was calculated as the closest geographical distance between each measurement location and one of the vertices in the U.S. Medium Shoreline dataset \cite{NOAAShore}, which includes all ocean and Great Lakes coasts for the contiguous 48 states. Because this dataset does not include the coastlines of Alaska and Hawaii, distance to coast calculations for these locations used manually extracted shorelines from NOAA's Shoreline Data Explorer \cite{NOAAShore2}. We acknowledge there are limitations to this method of distance calculation, as distances for some locations, such as Arizona (Flagstaff, Nogales, and Phoenix), are slightly longer than they would be had we used shoreline information for Mexico's Gulf of California. Nevertheless, these measurements effectively separate inland weather stations from coastal stations. 

Table \ref{tbl:variables} shows the weather variables included in our final analysis. We excluded mean daily measurements for temperature, precipitation, dew point, humidity and sea level pressure as these measurements were near perfect linear combinations of their corresponding minimum and maximum measurements. We also excluded maximum visibility from the analysis as this measurement was equal to 10 miles for more than 97\% of all recorded measurements. Lastly, we combined the information provided by maximum wind speed and maximum wind gust by retaining only the lower of the two measurements after removing outliers. The decision to combine the information from these two wind variables was motivated by the fact that 13\% of all maximum wind gust values were missing. In addition, it is difficult to separate unusually high, yet valid, maximum wind gust and wind speed measurements from true outliers.

Some stations did not record relevant climate variables. When possible, these missing observations were replaced with corresponding measurements obtained from the nearest National Weather Service (NWS) first order station as obtained through the National Climatic Data Center (NCDC) \cite{ncdc2018}. Missing values include wind speed in Baltimore, Maryland, precipitation in Denver, Colorado, and replacements of outlier precipitation measurements at multiple locations. When replacements were not readily obtained through the NCDC, systematic missing observations were replaced with corresponding observations from the nearest geographical neighbor within the dataset, as was the case for visibility and cloud cover in Baltimore, Maryland (replaced with Dover, Delaware, measurements) and Austin, Nevada (replaced with Reno, Nevada, measurements).

Table \ref{tbl:variables} also shows the observation ranges for each of the included variables. These measurement ranges are either definitional, such as the bounds for humidity, or simply practical, such as the bounds for temperature. All measurements falling outside the bounds shown in Table \ref{tbl:variables} were removed prior to our analysis. Several individual outliers were also removed or replaced based on location-specific inconsistencies including
\begin{itemize}
\item removal of one unusually low minimum temperature measurement in Honolulu, \linebreak[4] Hawaii, ($<10^\circ$F) and two in San Francisco, California ($< 20^\circ$F);
\item replacement of the following unusually high precipitation readings with precipitation readings at nearby weather stations \cite{ncdc2018}:
\begin{itemize}
\item Oklahoma City, Oklahoma, on 8/10/2017 ($38.33\mbox{in}  \rightarrow 0.8$in)
\item Salmon, Idaho, on 4/21/2015, 5/2/2016, and 5/3-4/2017 ($10.02\mbox{in} \rightarrow  0$in)
\item Flagstaff, Arizona, on 12/24/2016 ($7.48\mbox{in} \rightarrow  0.97$in)
\item Indianapolis, Indiana, on 7/15/2015 ($9.99\mbox{in} \rightarrow 0$in);
\end{itemize}
\item removal of one unusually low minimum dew point measurement in Honolulu, Hawaii ($<40^\circ$F), two in Hoquiam, Washington ($< 0^\circ$F), four in Las Vegas, Nevada \linebreak[4] ($<-15^\circ$F), and two in Denver, Colorado ($<-20^\circ$F).     
\end{itemize}

Forecast variables were restricted to minimum temperature, maximum temperature, and the probability of precipitation. We found no obvious outliers in the weather forecasts. This is reasonable due to the fact that forecasts are not subject to inevitable sensor technology failures that occur when taking an actual measurement. Rather, the forecast data were replete with duplicate values for minimum temperature and precipitation. We retained the lowest forecast of minimum temperature and the highest forecast of precipitation probability for each forecast. 

Forecast lags of six or seven days contained a large number of missing values. We removed all forecasts past lag 5. We also removed all forecasts containing negative lags (i.e., a forecast made \textit{after} the actual observation).

\section{Polar coordinate considerations for geographic maps \label{polar}}

The glyph plots in Figures \ref{fig:corGlyph} and \ref{fig:glyphPlot} rely on proper conversions from polar to geographic or Cartesian coordinates. This allows the glyphs to be plotted directly on the underlying map, rather than embedding polar coordinate subplots in the image. Avoiding subplots allows for greater precision in the placement of the glyphs and avoids the computational burden of creating and embedding multiple figures. This direct plotting approach requires special considerations for geographical maps, as polar coordinate glyphs become distorted when projecting geographical coordinates to a Cartesian plane. For example, a perfect circle in geographical coordinates will appear elongated in the vertical direction when the circle is projected in the northern hemisphere. One solution to this issue is to project all geographical coordinates to a Cartesian plane prior to the glyph construction. This can be conveniently accomplished using the mapproject() function in the mapproj R package \cite{mapproj}.

Polar coordinates are defined in terms of radius $r$ and angle $\theta$.  Figure \ref{fig:glyphPlot} defines $r \in [0, 1]$ as the scaled average absolute error between predicted and actual temperature and $\theta = \frac{(4-m)\pi}{6}$ where $m$ represents the numeric month. We center each glyph at 0 with Cartesian coordinates
\[
(x, y) = (r\cos{\theta}, r\sin{\theta})
\]
Let $(\mathbf{x}_i, \mathbf{y}_i)$ represent the set of Cartesian coordinates centered at the origin that create the glyph associated with location $i$. These coordinates are defined using the same units as the underlying map projection. The final coordinates of the rendered glyph are defined as
\[
\alpha\cdot(\mathbf{x}_i + u_x, \mathbf{y}_i + u_y)
\] 
where $(u_x, u_y)$ represents coordinates of location $i$ and $\alpha$ represents a global scaling parameter used to adjust the size of the rendered glyphs on the map. A point is drawn at location $(u_x, u_y)$ to serve as a reference for the glyph. Asymmetry about the point $(u_x, u_y)$ reveals seasonal patterns in the forecast errors. 

\begin{figure}
\centering
\includegraphics[width = 0.5\textwidth]{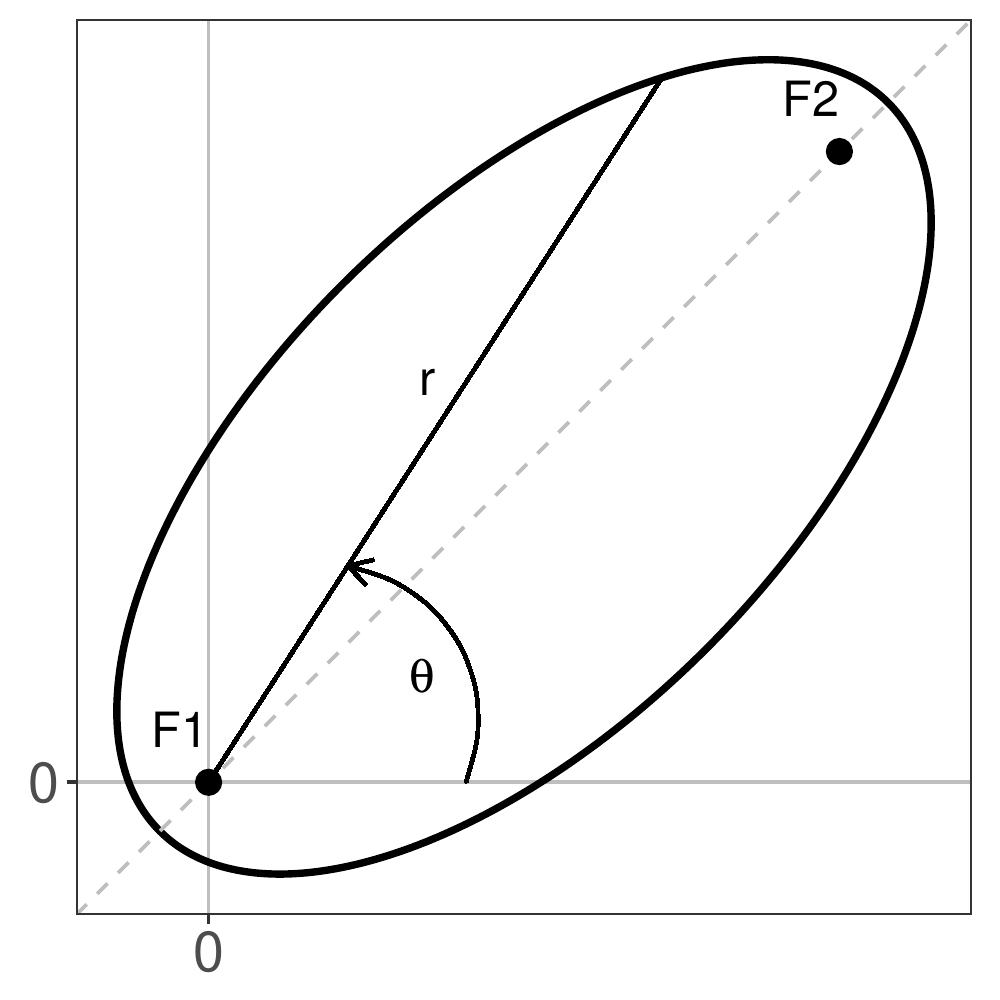}
\caption{Sample ellipse with $F_1$ located at the origin}
\label{fig:sampleEllipse}
\end{figure}

We construct the correlation ellipses of Figure \ref{fig:corGlyph} with foci $F_1, F_2$ located along the semi-major axis $y = x$ ($\theta = \frac{\pi}{4}$) for positive correlations and $y = -x$ ($\theta = -\frac{\pi}{4}$) for negative correlations. We fix $F_1$ at the origin and denote $r$ as the radius extending from $F_1$ to the edge of the ellipse, as illustrated in Figure \ref{fig:sampleEllipse}. This approach to ellipse creation is outlined in \citet{Knisley2001} and adapted here where we define $r$  for $\theta \in [0, 2\pi]$ as 
\[
r = \frac{(1-|\rho|)^2}{1-\sqrt{|p|(2-|p|)}\cos(\theta-\frac{\mbox{sign}(\rho)\pi}{4})},
\]
where $\rho \in (-1, 1) \backslash \{0\}$ represents the desired correlation between forecast errors. In the event that $\rho = -1, 1, \mbox{ or } 0$, we use $\rho \pm \epsilon\, (\epsilon > 0)$ when creating the ellipse to avoid numerical precision errors. The ellipse is then converted to Cartesian coordinates and centered at the origin as
\[
\left(r\cos(\theta) - \frac{|\rho|(2 - |\rho|)}{\sqrt{2}}, r\sin(\theta) - \mbox{sign}(\rho)\frac{|\rho|(2 - |\rho|)}{\sqrt{2}}\right).
\]
Each ellipse is scaled to be circumscribed in the $[-0.5, 0.5]\times [-0.5,0.5]$ square. This scaling makes it possible to create a matrix of ellipses using a common grid size.  It also reduces the difference in areas between ellipses which facilitates comparisons of shape. This scaling is defined as 
\[
(\mathbf{x}'_i, \mathbf{y}'_i) = \left(\frac{\mathbf{x}_i}{2\cdot\mbox{max}(|\mathbf{x}_i|)}, \frac{\mathbf{y}_i}{2\cdot\mbox{max}(|\mathbf{y}_i|)}\right).
\]

Note that there are three ellipses for each location. We define a matrix of ellipses centered at the shared vertex of the lattice denoted by $(u_x, u_y)$. Let $(\mathbf{x}_i, \mathbf{y}_i)$ represent the coordinates of one of the three ellipses centered at this location. Each ellipse is centered and scaled on the map as
\[
\alpha\cdot(\mathbf{x}'_i + u_x + o_{1}, \mathbf{y}'_i + u_y + o_{2})
\]
where $o_1$ and $o_2$ represent offset terms used to separate the centers of the three ellipses in the matrix defined for each location.  

This direct plotting approach of the ellipses eases plot customization, as there is no need to reconcile formatting differences between independently created subplots. This approach can also be generalized to plot other geometric shapes on a geographic map. It is also helpful for interactive applications that require fast renderings of images in response to dynamic inputs.  

\bibliographystyle{unsrtnat}  
\bibliography{dataExpo}

\end{document}